\documentclass[11pt,titlepage]{amsart}

\usepackage{graphics}
\usepackage{amssymb}
\usepackage{epsfig}
\usepackage[vcentermath]{youngtab}
\newcommand\bm\boldsymbol

\font\small=cmr8 scaled \magstep0
\outer\def\beginsection#1\par{\medbreak\bigskip
      \message{#1}\leftline{\bf#1}\nobreak\medskip
\vskip-\parskip
      \noindent}

\newcommand{\half}{\tfrac12}
\newcommand{\quart}{\tfrac14}
\newcommand{\ket}[1]{\big|#1\big\rangle}
\newcommand{\bra}[1]{\big\langle#1\big|}
\newcommand{\braket}[2]{\big\langle#1\big|#2\big\rangle}
\newcommand{\Tr}[1]{{\sf Tr}(#1)}
\newtheorem{thm}{Theorem}

\newcommand{\Mop}{\mathcal{M}}
\newcommand{\Hop}{\mathcal{H}}
\newcommand{\Pop}{\mathcal{P}}
\newcommand{\Nop}{\mathcal{N}}
\newcommand{\Oop}{\mathcal{O}}
\newcommand{\Sop}{\mathcal{S}}
\newtheorem{lemma}{Lemma}[section]

\newcommand{\oneone}{{\scriptstyle 11}}
\newcommand{\onetwo}{{\scriptstyle 21}}
\newcommand{\oneth}{{\scriptstyle 31}}
\newcommand{\onefo}{{\scriptstyle 41}}
\newcommand{\twoone}{{\scriptstyle 12}}
\newcommand{\twotwo}{{\scriptstyle 22}}
\newcommand{\thone}{{\scriptstyle 13}}

\newcommand{\reg}{$^{\text{\tiny\textregistered}}$}

\newcommand{\VW}{{\sf VW}}

\begin{document}

\titlepage
\begin{flushright}
\vspace{15mm}
{\sf UPRFT-06-11}
\end{flushright}
\vspace{15mm}
\begin{center}
  \huge{\sf The planar spectrum in  $U(N)$-invariant
    quantum mechanics by Fock space methods: I. The bosonic case.}

\vspace{15mm}

 \large{\bf R.\ De Pietri, S.\ Mori and E.\ Onofri\/}

   \vspace{5mm}
   {\small\sl Dipartimento di Fisica, Universit\`a di Parma}

 {\small\sl and}

 {\small\sl I.N.F.N., Sezione di Milano-Bicocca, Gruppo Collegato di Parma,
   43100 Parma, Italy}\\[1.5em]
\end{center}

\vskip 10mm
\centerline{\small\bf  Abstract}
\vskip 5mm
\noindent
\begin{flushleft}
  Prompted by recent results on Susy-U(N)-invariant quantum mechanics
  in the large N limit by Veneziano and Wosiek, we have examined the
  planar spectrum in the full Hilbert space of U(N)-invariant states
  built on the Fock vacuum by applying any U(N)-invariant combinations
  of creation-operators. We present results about 1) the
  supersymmetric model in the bosonic sector, 2) the standard quartic
  Hamiltonian. This latter is useful to check our techniques against
  the exact result of Brezin et al. The SuSy case is where Fock space
  methods prove to be the most efficient: it turns out that the
  problem is separable and the exact planar spectrum can be expressed
  in terms of the single-trace spectrum.  In the case of the
  anharmonic oscillator, on the other hand, the Fock space analysis is
  quite cumbersome due to the presence of large off-diagonal $O(N)$
  terms coupling subspaces with different number of traces; these
  terms should be absorbed before taking the planar limit and
  recovering the known planar spectrum. We give analytical and
  numerical evidence that good qualitative information on the spectrum
  can be obtained this way.
\end{flushleft}
\vfill
\begin{center}
{\sf\small [October, 4th 2006, revised November 10th, 2006]}
\end{center}

\newpage

\section{Introduction}
Interest in the planar (topological) expansion in QFT has always been
alive since the work of 't Hooft
\cite{hooft74:_planar_diagr_theor_for_stron_inter} and Veneziano
\cite{veneziano74:_large_n_expan_in_dual_model, veneziano76} (see also
\cite{ciafaloni75:_topol_expan_for_high}).  A recent revival of
interest has been triggered by a paper by Veneziano and Wosiek
\cite{veneziano06:_planar_quant_mechan} (hereafter \VW) followed by
\cite{veneziano06:_super_matrix_modelII,veneziano06:_super_matrix_modelIII},
about a class of supersymmetric quantum mechanical models, the
simplest among them being defined by
\begin{equation}
  \label{eq:1}
  H = \{Q,Q^\dagger\}, \; Q = \Tr{f^\dagger(a + g\, a^2)}
\end{equation}
where $f$ and $a$ are $N\times N$ operator valued Hermitian matrices,
with standard canonical (anti)\-com\-muta\-tion relations; the model
reveals a rich structure in the spectrum, in particular a peculiar
duality $b\rightarrow 1/b$, where $b^2=g^2N$ is 't Hooft's parameter.
The approach to the planar limit is based on the Fock space
representation of the Hamiltonian, which is for the time being the
only technique available in cases where a change of variable to
$U(N)$-invariant ``radial'' coordinates is not applicable
\cite{e.78:_planar_diagr,marchesini80:_planar_limit_for_su_n}. The
results presented by \VW\ are restricted to a peculiar subspace of the
full sector of $U(N)$ invariant states, namely the subspace
$\Hop^{(0)}$ spanned by the vectors which are obtained by applying a
``single trace operator'' to the Fock vacuum
\begin{equation}
  \label{eq:2}
 \Tr{ a^{\dagger n}}\, \ket{0}\,,\quad n=0,1,2,\ldots
\end{equation}

This subspace is considered ``dominant'', in the planar limit, with
respect to other states built over the vacuum by applying any product
of invariant operators. The Fock space approach to the large $N$ limit
was considered in \cite{thorn79:_fock_space_descr_of_n}, where the
issue of which kind of operators would leave the subspace of
single-trace states invariant in the planar limit was addressed.  The
\VW\ Hamiltonian (\ref{eq:1}) enjoys this property, while the
anharmonic oscillator Hamiltonian
\begin{equation}
  \label{eq:anharm}
  H = \half\Tr{p^2 + q^2} + \frac{g^2}{N}\Tr{q^4}
\end{equation}
does not. It follows that in this latter case it is necessary to
analyze the Hamiltonian in the full Hilbert space of $U(N)-$invariant
states. In the SuSy case it is clear that restricting to $\Hop^{(0)}$
is legitimate in the planar limit, but one misses in this way a big
portion of the spectrum, and, moreover, to estimate the $1/N$
corrections one has to study the full Hilbert space.

The multiplicity of states makes the calculation of the spectrum much
harder. To be specific, at a given number of quanta, i.e. considering
a product of $n$ creation operators on the vacuum, there exists one
state in $\Hop^{(0)}$ but there exist $p(n)$ states of the kind
\begin{equation}
  \label{eq:5}
  \Tr{ a^{\dagger \lambda_1}}\,  \Tr{ a^{\dagger  \lambda_2}}\,\ldots
 \Tr{ a^{\dagger  \lambda_m}}\,
  \ket{0}\,,\quad m\le n
\end{equation}
where $\lambda$ is any partition of $n$ and the function $p(n)$ counts
them all. It is known since Hardy and Ramanujan that $p(n)$ grows
exponentially ($p(n)=O(\exp\{K\surd n\})$, $K=\pi\sqrt{2/3}$ ), and
this is the main obstacle in a purely numerical approach. For
instance, while it is easy to diagonalize the \VW\  Hamiltonian
corresponding to a maximum number of quanta of thousands or even more,
the mere fact that $p(200)\approx 4\times 10^{12}$ (as it was first
computed exactly by MacMahon \cite{g.h.hardy02:_raman}) puts a severe
limit to the level one can explore numerically.

Our main result (Sec.2) will be the following: for the SuSy \VW\
Hamiltonian in the bosonic sector the Hilbert space of $U(N)$
invariant states splits into an infinite number of subspaces
$\Hop^{(m)}, m=1,2,3,...,$ which are left invariant by the Hamiltonian
in the planar limit. These subspaces are characterized by states with
a given number $m$ of trace factors in Eq.(\ref{eq:5}); so to speak
the number of traces is a good planar quantum number. As a
consequence, the Hamiltonian is recognized to be separable into the
sum of commuting operators, all unitarily equivalent to the single
trace operator studied by \VW. Essentially the Hamiltonian is
analogous to an isotropic multidimensional harmonic oscillator, where
the one-dimensional operator is given by \VW's operator. This fact has
several consequences, for instance if the 't Hooft parameter
$b^2=g^2N$ is grater than one, then all levels are infinitely degenerate.

The corrections of $O(1/N)$ break this degeneracy since they couple
the subspaces $H^{(m)}$ together. We do not have exact results on this
breaking mechanism, only numerical evidence. It appears that while
higher states' degeneracy is lifted, the first level (the zero mode)
keeps its infinite degeneracy even after the $1/N$ corrections are
taken into account.

For a second class of Hamiltonian operators, which typically do not
leave the Fock vacuum invariant, as is the case for the anharmonic
oscillator solved in \cite{e.78:_planar_diagr}, the analysis is more
involved. We give evidence on how the exact result, derived by
introducing ``radial variables'' for the $N$-dimensional matrix
$\bm x$, can be recovered numerically by reabsorbing the
off-diagonal $O(b^2 N)$ terms which deny the possibility to have
a straightforward planar limit. Our conclusion is that Fock space
methods are not particularly convenient for this kind of models, but
they can still be useful to explore the spectrum numerically in those
cases where a BIPZ technique is not applicable.

\section{Susy quantum mechanics in the U(N) invariant sector}

Let us introduce a bit of formalism. Let 
\begin{equation}
  \label{eq:3}
  \ket{\lambda} \equiv  \ket{\lambda_1,\lambda_2,\ldots,\lambda_m} ={\Nop}_\lambda \,\Tr{a^{\dagger \lambda_1}} \Tr{a^{\dagger \lambda_2}}\ldots \Tr{a^{\dagger \lambda_m}}\ket{0}\,.
\end{equation}
The right-hand side is symmetric in the exchange of $\lambda$'s, hence the left-hand side will always be reduced to a normal form subject to the condition
\begin{equation}
  \label{eq:6}
  \lambda_1\ge  \lambda_2\ge \ldots \ge \lambda_m\;.
\end{equation}
Where confusion may not arise, we shall denote the basis vectors
simply by $\ket{n,m,\lambda}$ or simply $\ket{\lambda}$ where
$\lambda$ is an $m$-partition of $n\in \mathbb{Z}^+$, a non-negative
integer; 
we write $|\lambda| = n$ or $\lambda\vdash n$ if $\lambda$ is a
partition of $n$ as it is customary in the mathematical literature.
The ``one-trace-states'' of
Ref.\cite{veneziano06:_planar_quant_mechan} are special cases, namely
$\ket{n,1,\{n\}}$.

If we refer our states to a basis of coherent states $
\ket{\bm z} $, such that
\begin{equation}
  \label{eq:7}
  a_{ij} \ket{\bm z} = \bar z_{ij}\ket{\bm z} 
\end{equation}
we realize that the states $\braket{\bm{z}}{n,m,\lambda}$ realize a
basis in the space of symmetric functions in the eigenvalues of $\bm
z$. Other basis, such as Schur's, can also be used where convenient
(see
e.g. \cite{ledermann77:_introd_to_group_charac,stanley99:_enumer_combin}): in
Sec.3 we shall turn to Schur's basis which coincides with the free
eigenbasis (for a detailed description of this formalism, see
\cite{stone90:_schur_funct_chiral_boson_and}).  One should take into
account the fact that our basis is not orthogonal. This is not going
to raise any problem, the only consequence being that the matrix
representing the Hamiltonian is not symmetric, but still unitarily
equivalent to a Hermitian one. It is however important to normalize
the states at least to leading order, to avoid having unnatural
dependence on $N$ in the matrix elements. So we decide to fix the
normalization factor to take into account the leading power in $N$:
\begin{equation}
  \label{eq:10}
  \Nop_\lambda= N^{-|\lambda|/2}\;.
\end{equation}
(this is just the scaling with $N$, the exact normalization is known,
but not actually necessary, see Appendix 2). Now we compute the matrix
elements of the Hamiltonian introduced in Eq.(\ref{eq:1}), as it is
reduced to the bosonic sector. For details about the algebra involved
the reader is referred to
\cite{VW2,veneziano06:_planar_quant_mechan}. We are led then to compute
\begin{equation}
  \label{eq:11}
  \left( \Tr{a^\dagger a} + g \Tr{a^{\dagger 2} a + a^{\dagger} a^2} + g^2 \Tr{a^{\dagger 2} a^2
    }\right)\; \ket{\lambda} = \sum_{\lambda'} \Hop_{\lambda',\lambda}\,\ket{\lambda'}\,.
\end{equation}
By applying the commutation relations 
\begin{equation}
  \label{eq:12}
  [a_{ij},a^\dagger_{kl}] = \delta_{il}\delta_{jk}
\end{equation}
one must bring all annihilation operators to the right eventually
reaching the vacuum. This is easily done symbolically using a language
like Mathematica\reg, but it is clear that the operation is going to
exhaust the capabilities of your computer as soon as the number
$|\lambda|$ reaches ten or a little more, because the multiplicity of
terms generated. A way out to this practical limitation is provided by
the following analytic result
(notice that the notation $\hat{\lambda}_j$ means that
 the term $\lambda_j$ must be deleted from the list):

\begin{thm}
\noindent
  \begin{enumerate}
  \item
\begin{equation*}
    \Tr{  a^{\dagger 2}\,a}\,\ket{\lambda_1,...,\lambda_m} = N^{1/2}\sum_j\,\lambda_j\,\ket{\lambda_1,...,\lambda_j+1,...,\lambda_m}
\end{equation*}
\item
\begin{eqnarray*}
  \Tr{  a^{\dagger 2}\,a^2}\,\ket{\lambda_1,...,\lambda_m} &=& \sum_{\substack{1\le j\le m\\\lambda_j>1}}\,\lambda_j\,
  \sum_{s=0}^{\lambda_j-2} \ket{\lambda_1,...,\lambda_j-s,...,\lambda_m,s}\\\nonumber
  &+& 2 \sum_{j < l}\,\lambda_j\,\lambda_l\,
  \ket{\lambda_1,...,\hat{\lambda}_j,...,\hat{\lambda}_l,...,\lambda_m,\lambda_j+\lambda_l}
\end{eqnarray*}
\item
\begin{eqnarray*}
  \Tr{  a^{\dagger}\,a^2}\,\ket{\lambda_1,...,\lambda_m} &=& N^{-1/2}\,\sum_{j}\,\lambda_j\,
  \sum_{s=1}^{\lambda_j-1} \ket{\lambda_1,...,\lambda_j-s,...,\lambda_m,s-1}\\\nonumber
  &+& 2\,N^{-1/2}\, \sum_{j < l}\,\lambda_j\,\lambda_l\,
  \ket{\lambda_1,...,\hat{\lambda}_j,...,\hat{\lambda}_l,...,\lambda_m,\lambda_j+\lambda_l-1}
\end{eqnarray*}
\end{enumerate}
\end{thm}
\begin{proof}
  \noindent \noindent The proof follows essentially the ideas
  introduced in \cite{thorn79:_fock_space_descr_of_n}: 

\noindent
$i)$ having
  just one annihilation operator to commute, the result is given by
  the sum of terms obtained by applying the commutator to each factor
  in $\ket{\lambda}$, and the result coincides with that one has for
  single trace states. The number of quanta on the right increases by
  1, hence the power of $N$ which takes into account the
  normalization. The resulting kets should be put in the standard form
  of Eq.(\ref{eq:6}) after the operation has been completed.

\noindent
$ii)$ The two annihilation operators can be made to
    commute with the same factor $a^{\dagger\lambda_j}$, provided
    $\lambda_j>1$, in which case we have a splitting into two traces
    or with two different factors, in which case we have a merge into
    one single trace. A factor 2 is due to the fact that the same pair
    $(j,l)$ can be coupled in two different ways to $a^2$. The term
    with $s=0$ produces a factor $N$ and this is the dominant term in
    the result. As before the ket arguments on the right-hand-side
    should be rearranged in order to have a non-increasing
    $m$-tuple. 


\noindent
$iii)$ As before, the two annihilation operators can be commuted with
the same $\Tr{a^{\dagger \lambda_j}}$ term, in which case we have a
splitting into two traces; when $s=1$ we get a factor $N$ which gives
the dominant contribution.  Otherwise the two $a$'s commute with two
distinct factors and we have a merge into a single trace.
  \end{proof}

  Notice that in case $(ii)$ the operator has a $g^2$ factor, which
  means that the dominant term $O(N)$ produces a finite result in the
  planar limit, the other being depressed by a factor $1/N$. In the
  other two cases the factor $g$ combines with the $N^{1/2}$ factor to
  give again a finite limit. Other terms provide off-diagonal
  corrections $O(1/N)$, which can be taken into account to get
  the planar expansion.

  Finally we have the matrix elements independent from $g=0$, that is
  $\Tr{a^\dagger\,a}\ket{n,m,\lambda}$; this is a diagonal operator,
  with eigenvalue $n=|\lambda|$, independently from the number of
  traces $m$.

  To get an idea of the full matrix, consider the portrait in Fig.1 (left),
  which includes all non-vanishing matrix elements, including the
  $O(1/N)$ terms, for $n\le 16$ (914 states)
  \begin{figure}[ht]
    \centering
    \includegraphics[width=.48\textwidth]{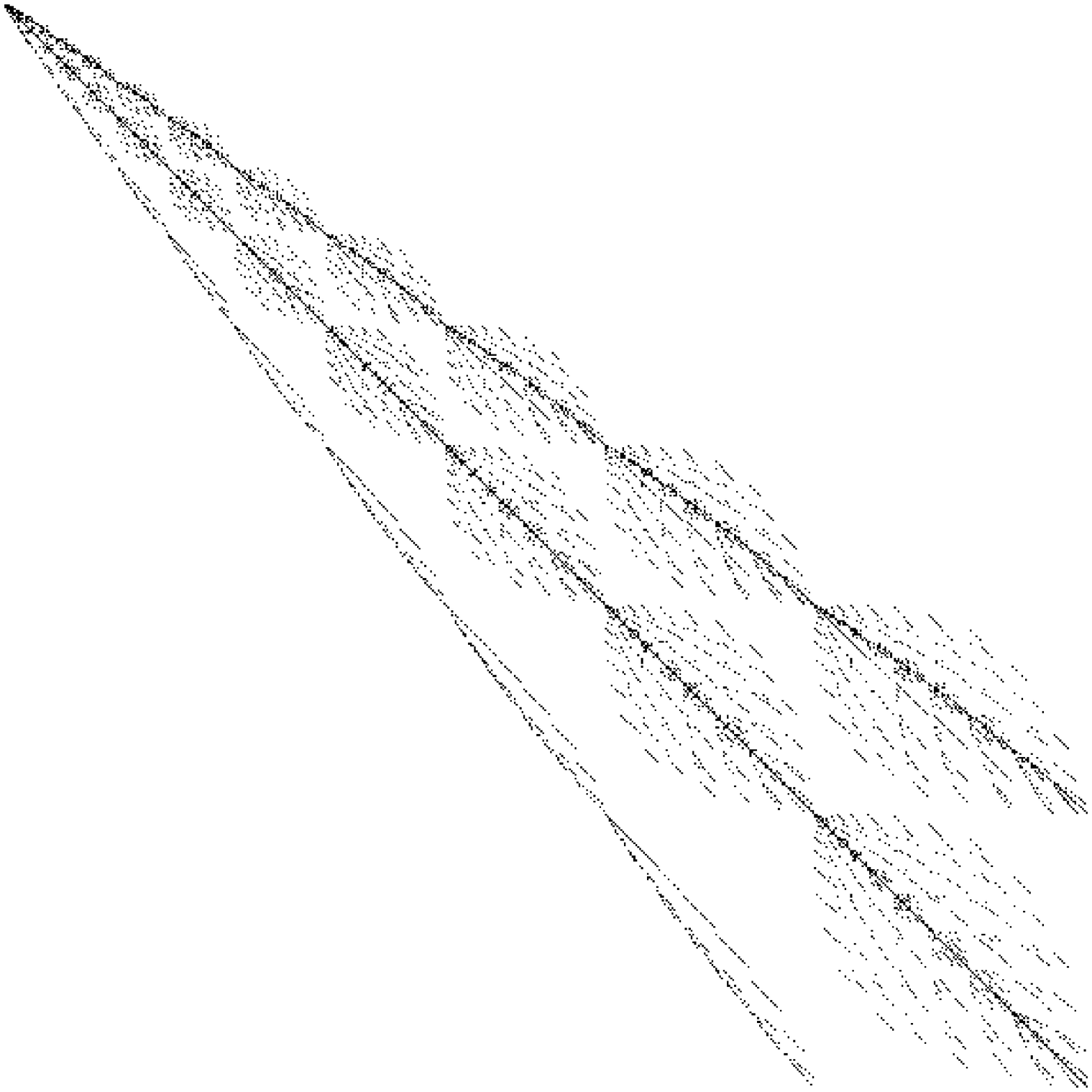}        
    \includegraphics[width=.48\textwidth]{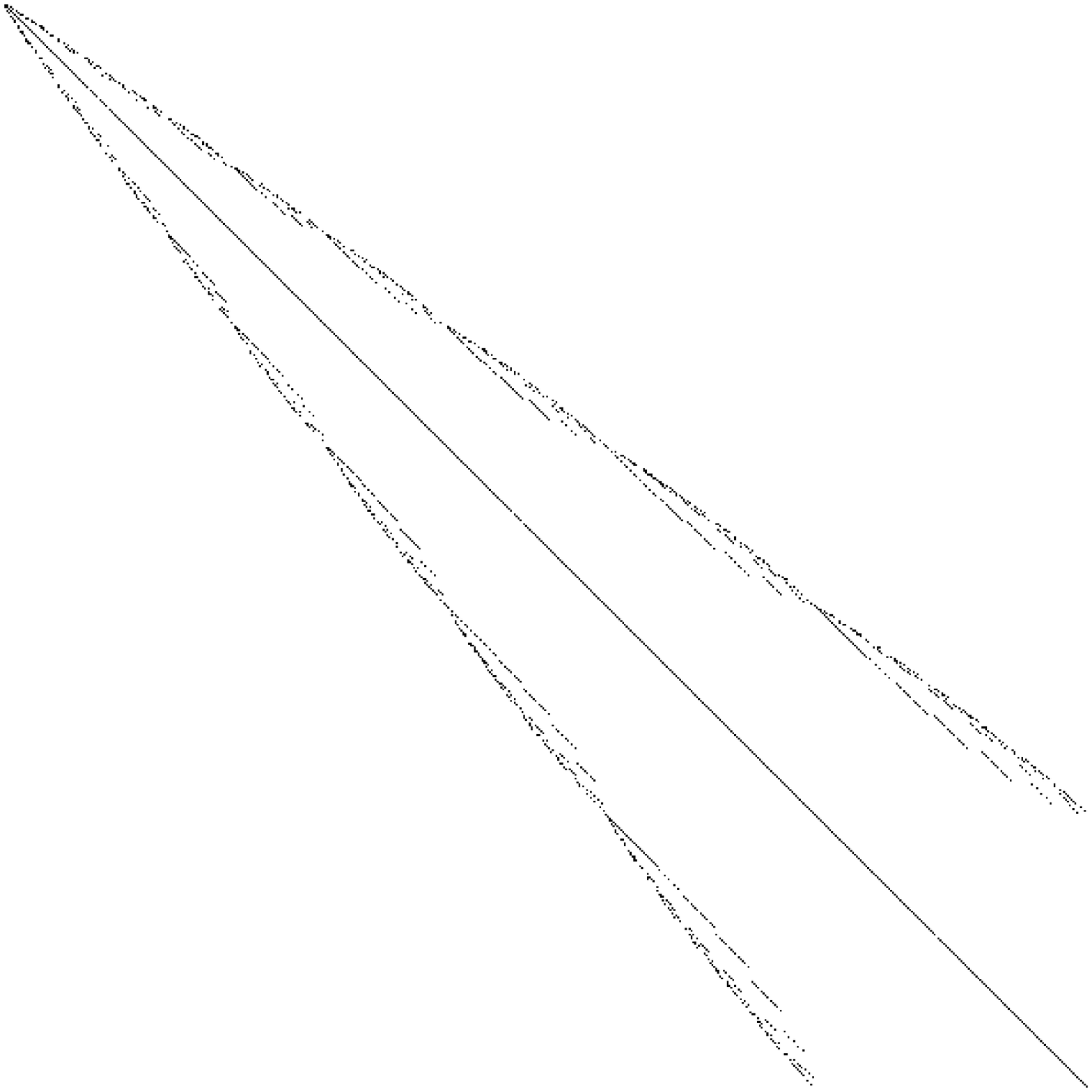}
    \caption{The sparsity pattern of $\Hop$ up to $\lambda\vdash 16$ (left);
      the same on the right after taking the planar limit.}
    \label{fig.1}
  \end{figure}
  Much simpler is the planar matrix, the finite part at $N=\infty$,
  which is depicted in Fig.1 on the right. 
  Both patterns are replicated with no substantial variations for
  higher values of $n$ (we checked up to $n=22$).

In spite of the apparent complexity of this matrix, which would
discourage from pushing to higher partition numbers - remember the
exponential growth of $p(n)$ - it turns out that the planar limit is
under full {\sl analytical\/} control. This is due to the following
observation: the finite matrix elements in the planar limit are all
generated when the Hamiltonian operator interacts with a single factor
$\Tr{a^{\dagger \lambda_j}}$, and the result is built by the sum of
all these contributions.  This means that in the planar limit the
number of traces in $\ket{\lambda}$ is unchanged by the application of
$H$, in other words the Hilbert space splits into invariant subspaces
$\Hop^{(m)}$ characterized by a fixed number $m$ of traces and
the planar Hamiltonian is given by the sum of commuting operators
$H_j$, each of them acting on the $j-th$ index in the
partition. Moreover each $H_j$ is identical to \VW\  operator in the
bosonic sector, that is the Hamiltonian restricted to single-trace
states. Hence we have
\begin{thm}
  The planar spectrum of the \VW\ SuSy operator in the full Hilbert
  space of $U(N)$ invariant states, as defined in Eq.~\eqref{eq:11},
  is given by the union of the spectra in each subspace $\Hop^{(m)},
  m=1,2,3,...$, namely
  \begin{equation}
    \label{eq:14}
    E_{n_1,...,n_m}=E^{(1)}_{n_1}+...+E^{(1)}_{n_m},\;n_1\ge n_2\ge...\ge n_m
  \end{equation}
  where $E^{(1)}_n$ are the eigenvalues of the Hamiltonian restricted
  to the single trace states as given in
  \cite{veneziano06:_planar_quant_mechan}.
\end{thm}
\begin{proof}
  The result follows from the property that the number of traces is left
  invariant in the planar limit; the constraint on the quantum
  numbers stems from the symmetry under permutation of the trace
  factors. (To make the paper more self contained, we give a derivation of
the \VW\  spectrum in Appendix 1).
\end{proof}
Notice that the block-diagonal structure of the planar matrix can be
made manifest if we rearrange the partitions in order of increasing
number of traces, as shown in Fig.2.

\begin{figure}[ht]
  \centering
  \includegraphics[width=.48\textwidth]{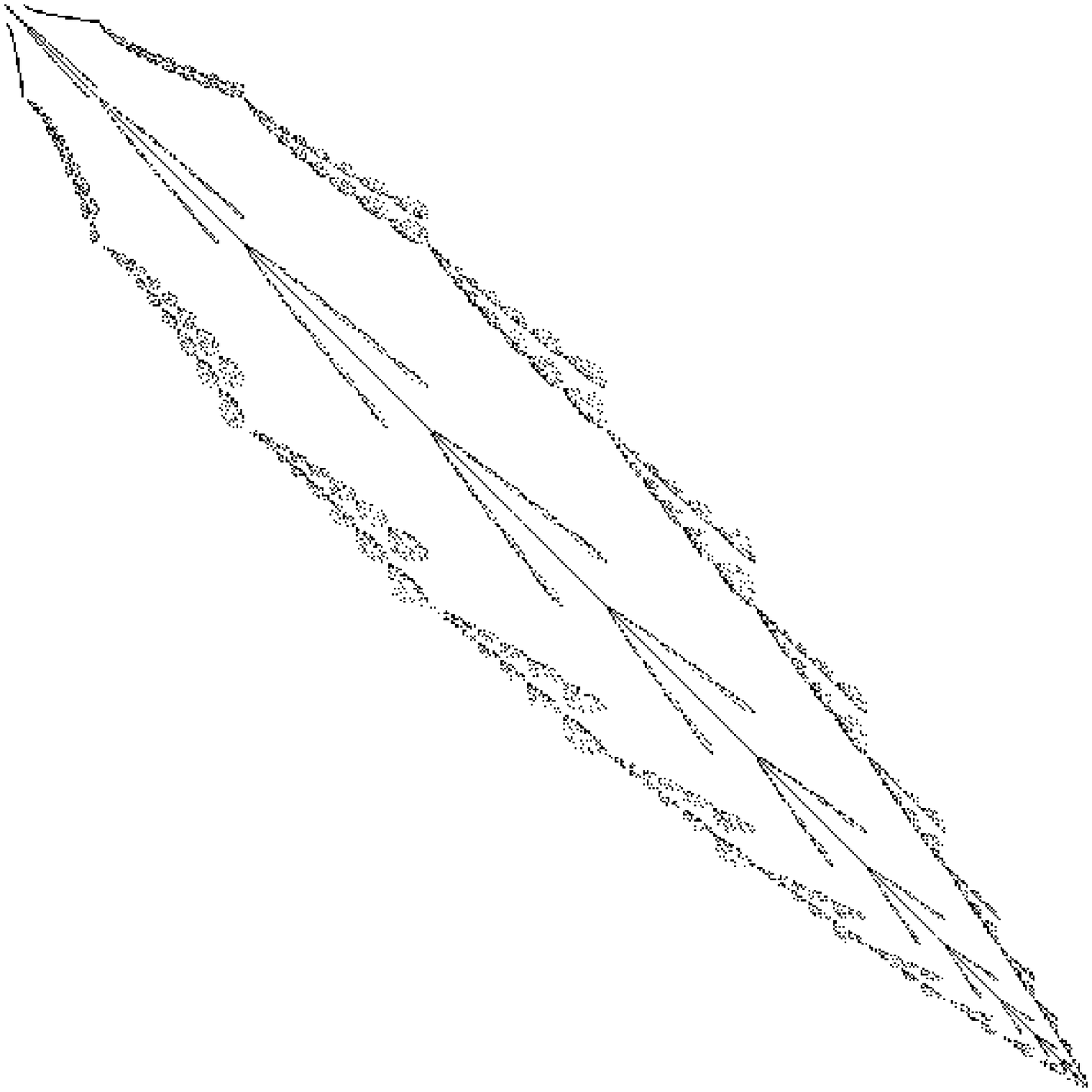}        
  \includegraphics[width=.48\textwidth]{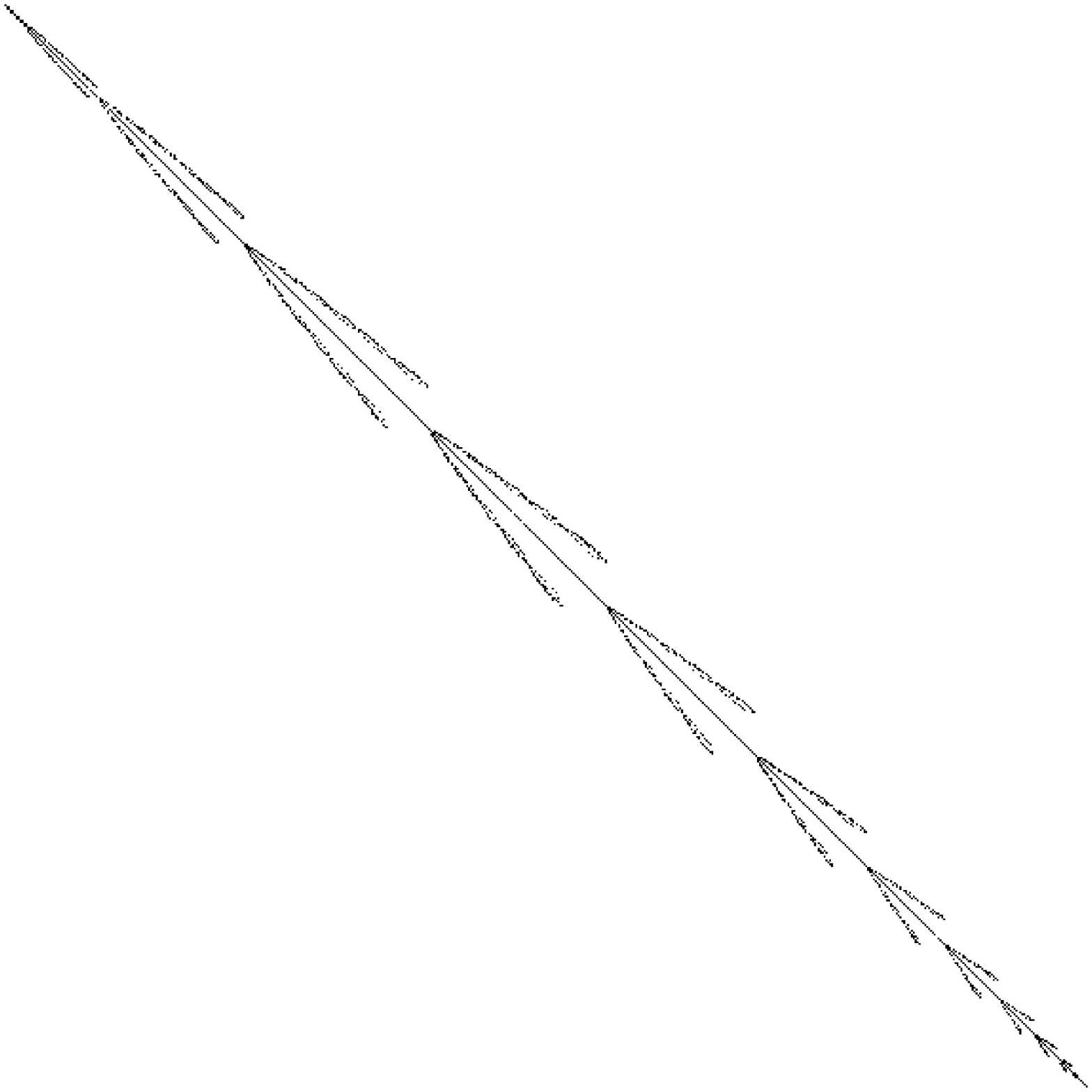}
  \caption{The rearrangement of $\Hop$ (left) and its block-diagonal
    structure in the planar limit (right).}
  \label{fig.2}
\end{figure}

We tested the result by diagonalizing each diagonal block.  In this
case we can explore larger matrices and the result is checked to high
accuracy.

Some comments on the structure of the spectrum are here in order.  In
the weak coupling limit the eigenvalues $E^{(1)}_n \approx (1+b^2) n,
(n>1)$, so the spectrum tends to be highly degenerate, like the
spectrum of the union of $m$-dimensional harmonic oscillators for
$m\ge 1$. Clearly, the states with $m$ traces set in only starting
with at least $m$ quanta, which implies that the multiplicity of
levels is not infinite, but it is steadily growing. See Fig.3 for the
case $b=1/4$, where the typical $p(n)$ multiplicity is manifest.

The spectrum is dramatically different at strong coupling. As it has
been shown in \cite{veneziano06:_planar_quant_mechan}, there is a
duality between weak and strong coupling, given by
\begin{equation}
  \label{eq:15}
  (E^{(1)}_n(b)+1)/b = b (E^{(1)}_{n+1}(1/b)+1)\,,\;(b^2=g^2N<1, n\ge 1)
\end{equation}
Moreover a new zero energy state arises at $b>1$,$
E^{(1)}_1(b)=0$. Since the \VW\ spectrum determines the whole planar
spectrum, the duality relation extends, with slight modifications, to
the general case. However the structure of the spectrum at strong
coupling is deeply changed, namely \emph{each eigenvalue is infinitely
  degenerate\/}. This is due to the fact that in the relation
(\ref{eq:14}) we may choose $n_2=n_3=...=n_m=0$ corresponding to the
zero modes and therefore there is an eigenvalue $E^{(1)}_n$ in each
invariant subspace $\Hop^{(m)}$. This fact extends to all other
eigenvalues: taking $n_3=n_4=...=n_m=0, m\ge 2$, one gets an infinite
number of eigenstates with the same eigenvalue
$E^{(1)}_{n_1}+E^{(1)}_{n_2}$ and so on.

This property can only be approximately displayed in a numerical plot,
since the diagonalization process implies a truncation on the
partition number, however a clear plateaus formation can be seen in
Fig.3 corresponding to $b>1$ (the plateaus would be infinitely long in
the exact solution).
\begin{figure}[ht]
  \centering
  \includegraphics[width=.48\textwidth]{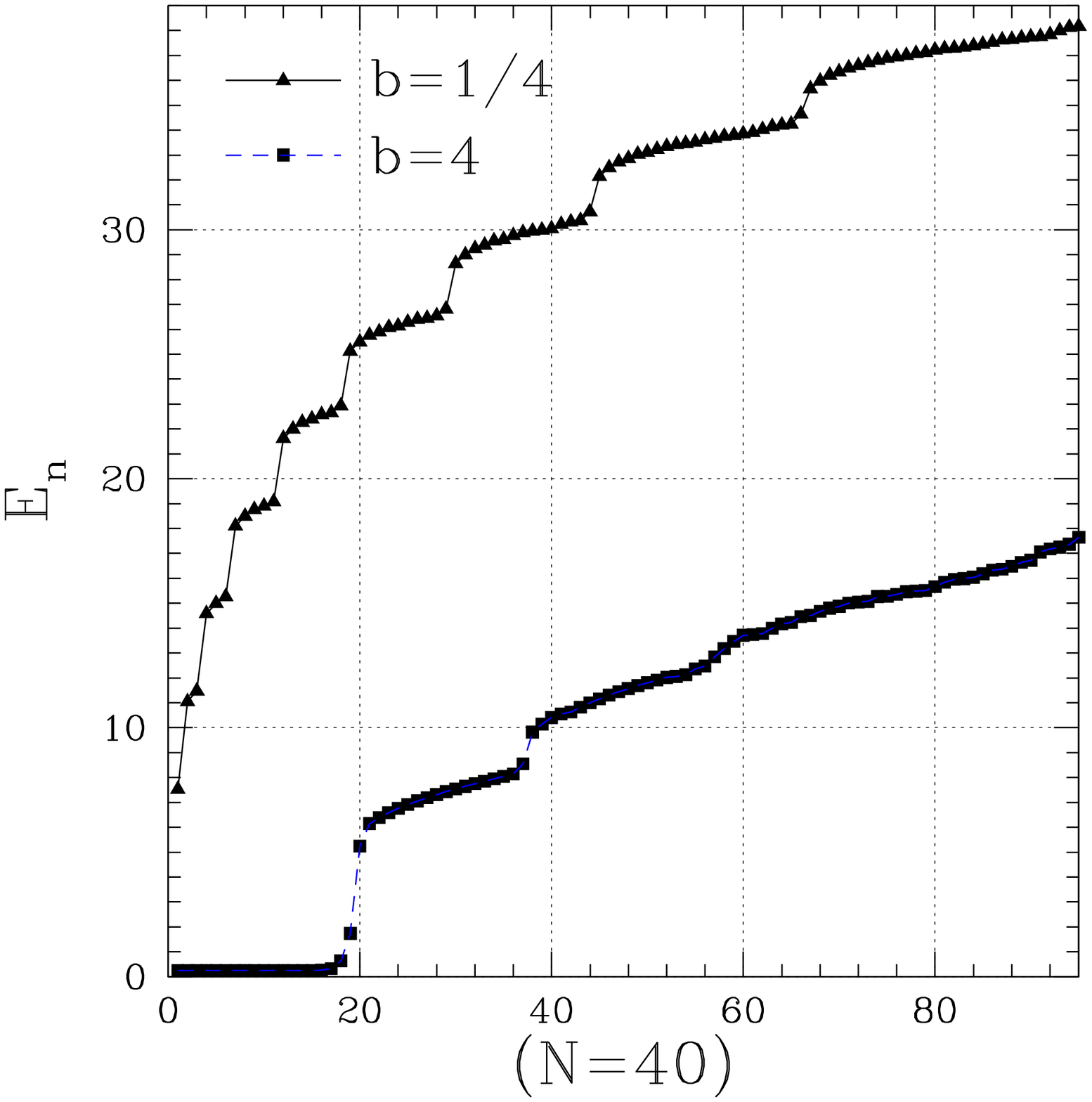}
  \includegraphics[width=.48\textwidth]{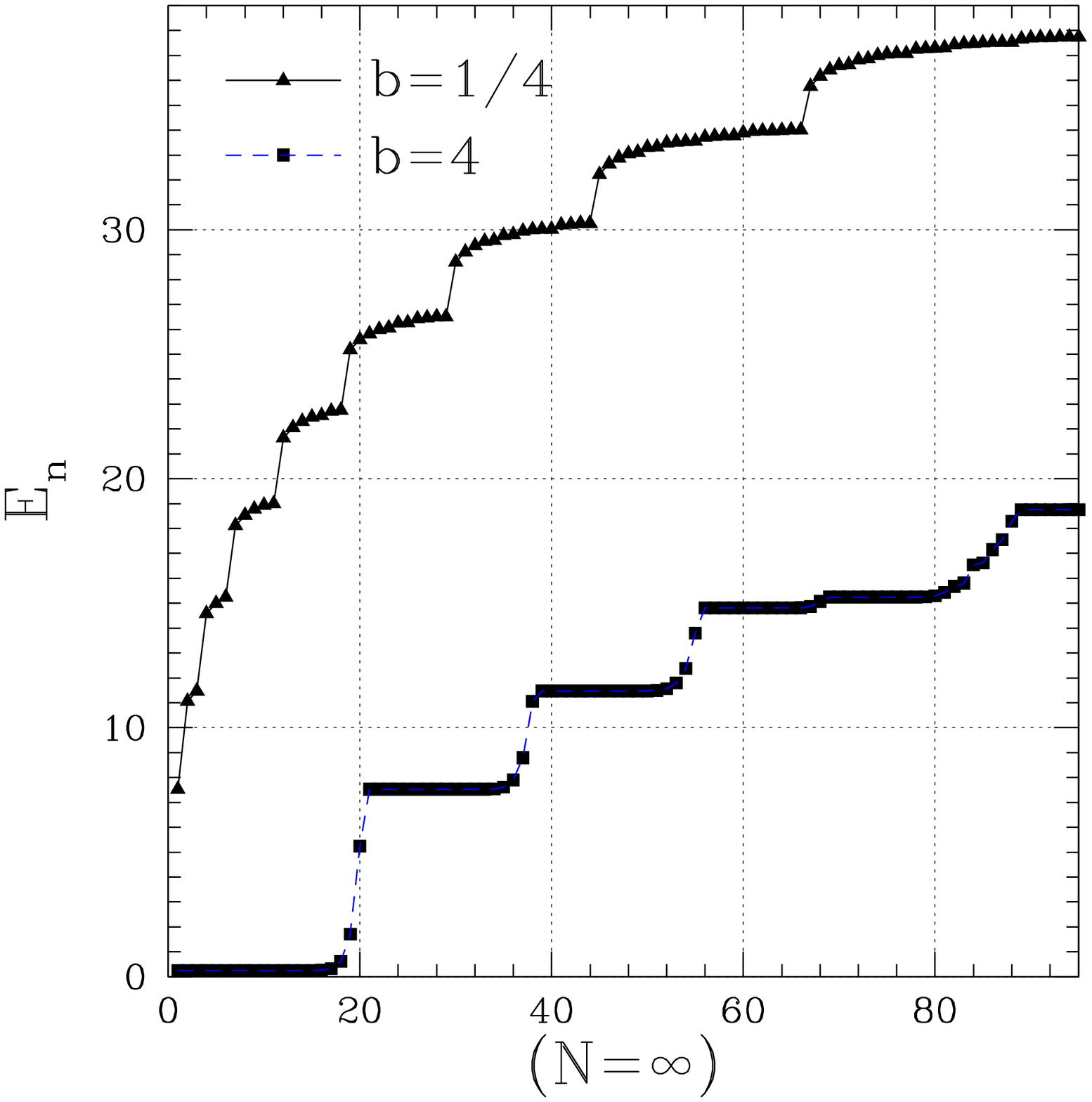} 
  \caption{The spectrum of $\Hop$, truncated at $\lambda\vdash 22$, for
    $N=40$ and $N=\infty$. }
  \label{fig:3}
\end{figure}

What happens of this infinite degeneracy at finite $N$? We do not have
analytic results on this matter. However diagonalizing the full
Hamiltonian including all subleading terms one may hope to get some
indications. In Fig.3 we report the two cases $N=40$ and $N=\infty$,
truncated at $\lambda\vdash 22$ (a $4507\times 4507$ sparse
matrix). We may observe that the degeneracy of higher levels tends to
be lifted by $O(1/N)$ corrections, while the ground state appears to
be robust against the corrections. We are lead to conjecture that the
zero modes are present even after switching on the subleading terms.

\section{The U(N) symmetric anharmonic oscillator}
The Hamiltonian of Eq.\eqref{eq:anharm} has been solved in the planar
limit in \cite{e.78:_planar_diagr,
  marchesini80:_planar_limit_for_su_n}. Here we want to analyze the
problem from the point of view of Fock space. The Hamiltonian deeply
differs from the SuSy model; for instance the Fock vacuum is not an
eigenstate, secondly, and more important, the Hilbert subspaces
$\Hop^{(m)}$ are {\sl not\/} invariant. These facts make the
Fock space approach much more involved than in the SuSy
case. Nevertheless we find it instructing how the exact result may
eventually emerge from the Fock space analysis. 

The Hamiltonian contains three main contributions (see
\cite{VW2} where a similar Hamiltonian is studied):
\begin{eqnarray}
  \label{eq:16}
  H &=& H_0 + H_2 + H_4\\\nonumber
H_0 &=& \half N^2 + \Tr{a^\dagger a} + \frac{b^2}{4N}\Tr{a^{\dagger 2}a^2 + ... + a^2a^{\dagger 2}}\\\nonumber
H_2 &=& \frac{b^2}{4N}\Tr{a^{\dagger 3}\,a + ... + a^3\,a^\dagger}\\\nonumber
H_4 &=& \frac{b^2}{4N}\Tr{a^{\dagger 4}+ a^4}
\end{eqnarray}
where the dots point to all other products generated by
$(a+a^\dagger)^4$, which differ from the terms displayed by the order
of the operators. It would be nice to have simple recipes like in the
previous section to compute matrix elements. In this case however the
problem is much more involved, with 16 quartic operators, some of them
containing two annihilation operators or more. To get an idea of the
work required, just consider matrix elements between states with
partition number (occupation number) ten, that is a $42\times42$ block
on the diagonal: this makes 10584 matrix elements. If we have to reach
a matrix truncation corresponding to a maximum $|\lambda|$ of order at
least 20, the total number of matrix elements to be computed touches the $10^8$
range, clearly unfeasible by brute force. But also an approach through
a computer algebra system is going to face serious difficulties if we
want to raise the truncation number above 10, because of the
proliferation of terms in the reduction of operator products. So we
need some kind of human-assisted-computer-algebra-approach.

\subsection{Matrix elements by computer algebra at low $|\lambda|$ }
We have set up a Mathematica code which is able to compute the matrix
elements by blindly applying the commutation relations. Let us observe
that since we initially work with the basis of Eq.(\ref{eq:3}) which
is not orthogonal, the matrix made of scalar products $
\bra{\lambda}\, H \, \ket{\mu}$ must be multiplied on the left by the
inverse of the metric matrix $ \Mop_{\lambda\mu} =
\braket{\lambda}{\mu} $ before we can use it to compute the
spectrum. This is the true representative matrix that we denote by
$\Hop$
\begin{equation}
  \label{eq:17}
  \Hop_{\lambda\mu}=\Mop^{-1}_{\lambda\nu}\bra{\nu}H\ket{\mu}
\end{equation}  
We start examining the main block on the diagonal of the Hamiltonian
in the ``partitions basis'' $\ket{\lambda}$. From the analysis of low
dimensional blocks a regular pattern emerges. After subtracting a term
$\half N^2(1+b^2)$ proportional to the identity operator, we are left
with an ``almost'' diagonal matrix, in the sense that the off-diagonal
matrix elements are depressed by a factor $1/N$. For instance:
{\small \begin{eqnarray*}
    \Hop_{00} &=&  \quart b^2\\
    \Hop_{11} &=& \quart b^2 + 1 + 3 b^2\\
    \Hop_{\lambda\vdash 2,\mu\vdash 2} &=&
  \begin{pmatrix}
     \quart b^2 + 2(1 + 3 b^2) & \frac{3b^2}{N}\\
\frac{3b^2}{N}& \quart b^2 + 2(1 + 3 b^2)
  \end{pmatrix}\\
 \Hop_{\lambda\vdash 3,\mu\vdash 3} &=&
  \begin{pmatrix}
       \quart b^2 + 3(1 + 3 b^2) & \frac{6b^2}{N} & 0\\
\frac{9b^2}{N} &  \quart b^2+ 3(1 + 3 b^2)&\frac{ 9b^2}{N}\\
0 & \frac{3b^2}{N} &  \quart b^2+ 3(1 + 3 b^2)
  \end{pmatrix}\\
 \Hop_{\lambda\vdash4,\mu\vdash 4} &=&
  \begin{pmatrix}
       \quart b^2 + 4(1 + 3 b^2) &\frac{9b^2}{N} & \frac{12b^2}{N}&0 &0\\
\frac{12b^2}{N}&  \quart b^2+ 4(1 + 3 b^2)&0&\frac{ 12b^2}{N}&0\\
\frac{6b^2}{N}& 0 & \quart b^2+ 4(1 + 3 b^2)&\frac{ 3b^2}{N}& 0\\
0&\frac{9b^2}{N}& \frac{6b^2}{N}&  \quart b^2+ 4(1 + 3 b^2)&\frac{18b^2}{N}\\
0&0&0&\frac{3b^2}{N}&  \quart b^2+ 4(1 + 3 b^2)
  \end{pmatrix}\\
\end{eqnarray*}
} 

A part from the obvious occurrence of a multiple of the identity of
dimension $p(n)$ with a value $n(1+b^2) + \quart b^2$, which obviously
generalizes to all higher dimensional blocks, the rest of the matrix
is not readily generalizable, at first sight. 

The off-diagonal blocks are generated by $H_2$ and $H_4$. Let's
examine the first few of them to see the pattern (we discard terms
$O(1/N^2)$):
\begin{eqnarray*}
 \Hop_{\lambda\vdash 2,\mu\vdash 0} &=&
  \begin{pmatrix}
Nb^2\\
\half b^2
  \end{pmatrix}\\
 \Hop_{\lambda\vdash 3,\mu\vdash 1} &=&
  \begin{pmatrix}
b^2\\ Nb^2\\\half b^2
  \end{pmatrix}\\
 \Hop_{\lambda\vdash4,\mu\vdash2} &=&
  \begin{pmatrix}
2b^2&0\\
0&2b^2\\
Nb^2&0\\
\half b^2&Nb^2\\
0&\half b^2
  \end{pmatrix}\\
\end{eqnarray*}
\begin{eqnarray*}
 \Hop_{\lambda\vdash0,\mu\vdash2} &=&
  \begin{pmatrix}
    2b^2N\!+\!\frac{b^2}{N}\,&\,3b^2
  \end{pmatrix}\\
 \Hop_{\lambda\vdash1,\mu\vdash3} &=&
  \begin{pmatrix}
9b^2\,&\,2b^2N\!+\!13\frac{b^2}{N}\,&\,9b^2
  \end{pmatrix}\\
 \Hop_{\lambda\vdash2,\mu\vdash4} &=&
  \begin{pmatrix}
12b^2\,&\, 18\frac{b^2}{N}\,&\,4b^2N\!+\!26\frac{b^2}{N}\,&\,3b^2\,&\,0\\
12\frac{b^2}{N}\,&\,b^2\,&\,0\,&\,2b^2N\!+\!25\frac{b^2}{N}\,&\,18b^2
  \end{pmatrix}\\
\end{eqnarray*} 

The matrix elements for $H_4$ are similar in that they contain
``superleading'' terms $O(b^2N)$ which do not allow for a
straightforward planar limit. Unlike the $O(N^2)$ along the diagonal
which only contributes to the zero-point energy, the big off-diagonal
terms cannot be disposed of in a simple way and they represent the
main obstacle to compute the planar limit. With hindsight from
\cite{e.78:_planar_diagr}, we know that the zero point energy has a
full expansion in powers of $b^2$, which means that contributions
coming from the off-diagonal terms are needed to recover the exact
result. We shall postpone the discussion on this point after we switch
to another basis, that of Schur's functions.

\subsection{The basis of Schur's functions}
It is well-known from the theory of symmetric functions (see e.g
\cite{ledermann77:_introd_to_group_charac,stanley99:_enumer_combin})
that there are many interesting basis in the ring of symmetric
functions on $N$ indeterminates, $\Lambda(N)$. Up to now we worked
with the ``partitions basis'' of Eq.(\ref{eq:3}); it's easy to relate
this basis to the corresponding basis in $\Lambda(N)$. Any vector
$\lambda$ can be represented in the basis of coherent states where it
becomes a symmetric polynomial of the eigenvalues
$\zeta_1,...,\zeta_N$ of the matrix $\bm{z}$
\begin{equation}
  \label{eq:18}
  \braket{z}{\lambda} = \Tr{z^{\lambda_1}}... \Tr{z^{\lambda_m}} = \sum\zeta_j^{\lambda_1}\,\sum\zeta_j^{\lambda_2}\,...,\sum\zeta_j^{\lambda_m}\;.
\end{equation}

It is now possible to consider other basis in $\Lambda$ to explore the
properties of $H$. One such basis is defined by Schur's functions (see
\cite{stone90:_schur_funct_chiral_boson_and} for an application). This
task can be easily achieved by symbolic computer algebra. Let us
denote by $\Hop^\Sop$ the representation of $H$ in Schur's basis.  The
main feature of $\Hop^\Sop$ turns out to be the following: all blocks
along the diagonal are now in {\sl exact diagonal form}; the diagonal
matrix elements  in the $n-$th block are now easily identified, being given by the
already computed $n(1\!+\!3b^2)\!+\!\quart b^2$ plus corrections
$O(1/N)$ which follow a regular pattern, e.g. for $1\le n\le 6$ they
are given by

\begin{table}[ht]
   \begin{center}
     \begin{tabular}{|l||c|}\hline
$n$      & \mbox{$\delta E=b^2/N\,\times$}\\\hline
0&0\\\hline
1&0\\\hline
2&(3,-3)\\\hline
3&(9,~0,-9)\\\hline
4&(18,~6,~0,-6,-18)\\\hline
5&(30,15,~6,~0,-6,-15,-30)\\\hline
6&(45,27,15,~9,~9,~0,-9,-9,-15,-27,-45)\\\hline
\end{tabular}\label{tab:1}
\end{center}\caption{The fine structure of anharmonic oscillator to
  first order in $1/N$.}
\end{table}

\noindent
The pattern may not be clear at first sight, but it emerges
immediately if we consider the partitions and Ferrers-Young diagrams
associated to them, as it is familiar from the theory of the symmetric group;
namely to each partition $\lambda_1\ge\lambda_2\ge...\ge \lambda_m$ it
is useful to associate a diagram composed of $m$ lines each
containing $\lambda_k$ square cells, i.e. 
\begin{equation*}
  \{5,2,1\}\rightarrow \yng(5,2,1)
\end{equation*}

 The formula reproducing the data of Tab.~1 is extremely simple
\begin{equation}
  \label{eq:19}
  \delta E_\lambda = 3\sum_{(i,j)\in\lambda}(i-j)\,b^2/N.  
\end{equation}
where the pair $(i,j)$ runs on the cells of the Ferrers-Young diagram
of the partition ($i$ is the column and $j$ the row index,
respectively).  For instance at $n=6$ we find
\begin{equation*}
  \begin{split}
  \label{eq:20}
\lambda=\{6\}\rightarrow \yng(6)\hfill\rightarrow
3\times\sum  \begin{bmatrix}
    0&1&2&3&4&5
  \end{bmatrix}
& \rightarrow\hfill 45\\
\lambda=\{5,1\}\rightarrow  \yng(5,1)\hfill\rightarrow
3\times\sum   \begin{bmatrix}
    ~0&1&2&3&4\\
    -1&&&&
  \end{bmatrix}
&  \rightarrow\hfill 27\\
  \ldots
  \end{split}
\end{equation*}
By construction the correction is opposite in sign for conjugate
partitions, and it averages to zero on every diagonal block, a feature
which was already noticed in \cite{marchesini80:_planar_limit_for_su_n}.

Also off-diagonal blocks greatly simplify in Schur's basis. First of
all, since the basis {\sl is\/} orthogonal, $\Hop^\Sop$
is symmetric, hence we save half of the computing effort. Secondly,
the superleading terms turn out to be very regular: all coefficients
of $b^2N$ are either $1$ or $-1$, for instance 
\begin{equation*}
  \Hop^\Sop_{\lambda\vdash3,\mu\vdash5}= b^2N
  \begin{pmatrix}
    ~1&~0&~1&-1&~0&~0&~0\\
~0&~1&~0&~0&~0&-1&~0\\
~0&~0&~0&~1&-1&~0&-1
  \end{pmatrix}+\text{{\sl finite terms}}
\end{equation*}

\subsection{The role of superleading terms}
We now want to discuss how superleading terms of $O(b^2N)$ contribute
to the planar expansion and allow in principle to recover the known
result. At this stage we have an expansion for the matrix representing
the Hamiltonian in Schur's basis of the following form
\begin{equation}
  \label{eq:8}
   \Hop^\Sop = \half N^2(1\!+\!b^2)\mathbb{I} + N\, T_1 + T_0 + N^{-1} T_{-1} +...
\end{equation}

The term $T_1$ should be taken into account {\sl before\/} taking the
limit $N\to \infty$. It is clear that off-diagonal terms of order $N$
will contribute corrections of order $N^2$ to the eigenvalues, however
there should not be any higher order corrections, since there are no
such terms in the planar expansion; moreover these corrections should
be the same for all eigenvalues from what we know on the basis of the
exact solution. The effect of $T_1$ should then reduce to a mere
redefinition of the vacuum energy.
We can deduce even more: $T_1$ cannot have by itself any discrete
eigenvalue $t_1$, otherwise, in the strong coupling limit, we would
find, by perturbation theory in $1/b$, a leading behavior $b^2N t_1$
which is absent in the exact solution. Hence we are led to argue that
$T_1$ must be an operator with continuous spectrum which perturbs
$T_0$ by simply shifting the zero energy level, but leaving the energy
gaps unchanged. An example of this kind comes from elementary quantum
mechanics. Consider the simple harmonic oscillator perturbed by a term
$-F x$; its perturbative expansion stops at second order for all
levels and its only effect is to shift all the spectrum by a fixed
amount $-\half F^2$. We argue that this is happening with $T_1$ and
this can be proven by a direct analysis showing that $T_1$ is
unitarily equivalent to a component of the position operator in an
anisotropic two-dimensional harmonic oscillator (see Appendix 2).  As
in the harmonic oscillator example, it would be natural to change the
vacuum state in order to re-absorb the term $T_1$. This will not be
attempted here and is left for future investigations\footnote{This
  idea was formulated by G.Veneziano, private communication.}.

Let us observe that the general picture we get at this stage, a ground
state proportional to $\epsilon(b)\,N^2$ and a splitting of order
$1/N$ given by Eq.~(\ref{eq:19}) is qualitatively correct, and this 
result is due to the introduction of a convenient basis, provided by Schur's
functions.

\subsection{The Hamiltonian matrix from group theory}
Even if Schur's basis allows for a better starting point in attacking
the problem, still the limitations posed by the brute force symbolic
calculation do not allow us to reach a reasonable truncation of the
Hamiltonian. But there is another way in which the theory of symmetric
functions, or in other words the theory of the symmetric group, can
help in the calculation of matrix elements. Let us denote by $
\ket{\lambda}_{\Sop}$ the vector corresponding to the Schur's function
\begin{equation*}
  s_\lambda(x) = \frac{\det(x_j^{\lambda_i+i-1})}{\det(x_j^{i-1})}
\end{equation*}
while we keep the symbol $\ket{\lambda}$ for the states constructed
via application of multiple traces (Eq.(\ref{eq:3})). The
transformation between the two basis
\begin{equation}
  \label{eq:9}
  \ket{\lambda}_{\Sop} = \sum_\mu S_{\lambda\mu}\ket{\mu}
\end{equation}
is well-known and is directly related to the character tables of
$S_n$, the group of permutations (see
\cite{ledermann77:_introd_to_group_charac}, Ch. 4):
\begin{equation*}
  S_{\lambda\mu}=\frac{1}{n!}d_\mu\,\chi_\mu^\lambda
\end{equation*}
where $n=|\lambda|=|\mu|$, $d_\mu$ is the dimension of the irreducible
representation indexed by $\mu$ and $\chi$ is its character. We show
now how the knowledge of $S$ can simplify
the computation of the matrix elements. Let us compute
\begin{equation*}
  \begin{split}
    \Tr{a^{\dagger 4}}\ket{\lambda}_{\Sop} &=
   \sum_\mu\,S_{\lambda\mu} \Tr{a^{\dagger 4}}\ket{\mu}=\\
    &\sum_\mu\,S_{\lambda\mu} \ket{\mu'}\bigg\vert_{\mu'=\mu\cup\{4\}}=\\
    &\sum_{\mu,\nu}\,S_{\lambda\mu}S^{-1}_{\mu'\,\nu} \ket{\nu}_{\Sop}
  \end{split}
\end{equation*}
(here $\lambda\cup\mu$ denotes the standard set-theoretic 
{\sl union\/} of two lists).  The point is that the character table can be
easily built to rather large dimensions using the results of Jacobi
and contemporaries and this allows to break the limit of the brute
force calculation.  The normalization factor is easily found to be
\begin{equation*}
_{\Sop}\braket{\lambda}{\mu}_{\Sop}=\delta_{\lambda\mu}
\prod_{(i,j)\in\lambda}\left(N+i-j\right)
\end{equation*}
where the indices $i,j$ run on the cells of the Ferrers-Young diagram
of the partition ($i$ is the column and $j$ the row index,
respectively), e.g.

\begin{equation*}
  \{4,2,1\}\to\young(\oneone\onetwo\oneth\onefo,\twoone\twotwo,\thone)
  \to
  \begin{bmatrix}
    ~0&~1&~2&~3\\
    -1&~0&&\\
    -2&&&\\
  \end{bmatrix}
\end{equation*}
which gives the normalization factor
\begin{equation*}
_{\Sop}\braket{\{4,2,1\}}{\{4,2,1\}}_{\Sop}=
N(N+1)(N+2)(N+3)(N-1)N(N-2)
\end{equation*}

A similar formula can be found also for the matrix elements  of  $H_2$ in the
Hamiltonian containing three creation and two annihilation
operators. In this way we have built the matrix in the (orthonormal)
Schur's basis up to $\lambda\vdash 21$ and presumably this may be
improved. The matrix can be used to explore the spectrum in the planar
limit by expanding in powers of $1/N$ and neglecting terms $O(1/N^2)$
or smaller. In particular we can study the crucial question about the
ground state, which should be compared with the exact result (see
Fig.4 where the straight line corresponds to $\half N^2(1+b^2)$).
\begin{figure}[ht]
  \centering
\includegraphics[width=12.cm,height=7.cm]{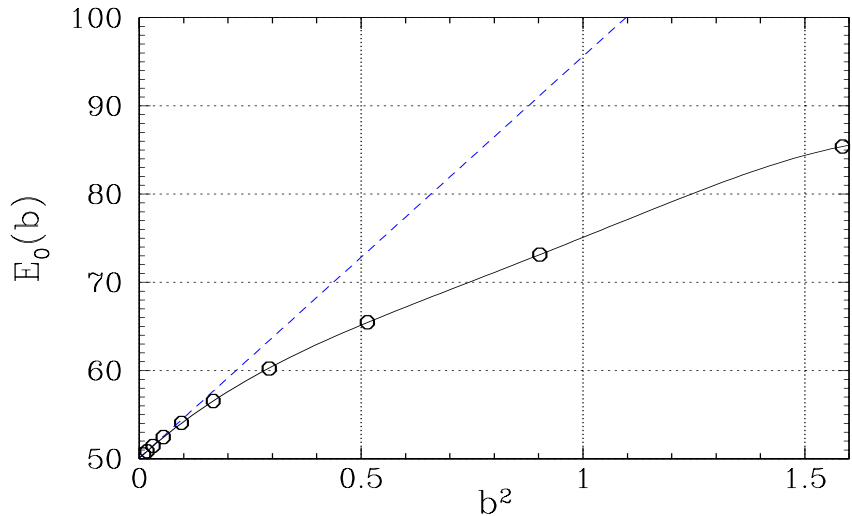}
  \caption{The ground state energy as a function of $b^2$ at $N=10$.}
  \label{fig:4}
\end{figure}
As a final comment, the introduction of Schur's basis appears the most
natural choice in view of the structure of the exact solution, a
collection of $N$ non-interacting fermions (see
\cite{e.78:_planar_diagr}).  Schur's functions are also known as
Slater-Fock determinants, of course.

\section*{Conclusions}
Fock space methods are very efficient for a class of $U(N)$-invariant
Hamiltonians similar to \VW\ model, where the Fock vacuum is
stable. We derived the exact spectrum for the SuSy model in the
bosonic sector, and this should be easily extended to the one-fermion
sector. The one-trace sector, while representing a small subspace of
the whole Hilbert space, provides the building block for understanding
the whole spectrum. The $b^2=1$ transition marks a boundary between a
typical approximate $p(n)$ degeneracy to a regime with infinite
degeneracy, which is in general broken by $O(1/N)$ corrections which
we have identified. Extension of these techniques to non-vanishing
fermion number will be considered in a future publication.  For the
anharmonic oscillator, as well as for all other Hamiltonians covered
by the exact BIPZ solution in the planar limit, Fock space methods are
not convenient. Still, by a combined effort of analytical and computer
algebra techniques, they may prove to be reliable in those cases where
an approach based on ``polar coordinates'' \`a la BIPZ is not
available. Our results show that superleading matrix elements $O(b^2
N)$ do not spoil the planar expansion, and the spectrum can be at
least computed numerically \footnote{Mathematica codes used in this
  work will be made available to interested readers, on request.}.

\section*{Acknowledgments}
Several interesting discussions with G. Veneziano and J. Wosiek are
gratefully acknowledged.  We warmly thank G. Cicuta for his generous
advice and fruitful collaboration while this work was conducted.  We
thank the Referee for many useful suggestions which helped in
improving the final version of the paper. 

\section*{Appendix 1}
The spectrum of the SuSy model in the single trace subspace has been
computed exactly in \cite{veneziano06:_planar_quant_mechan}. We
give here an equivalent derivation, essentially algebraic in
character. 

The matrix to be diagonalized is the following
\begin{equation*}
  \bra{n}\Hop\ket{m}=(1+b^2(1-\delta_{n1}))\,n\,\delta_{nm}+
  b\,\sqrt{m(m+1)}\delta_{n,m+1}+  b\,\sqrt{n(n+1)}\delta_{m,n+1}\,.
\end{equation*}
Another matrix, differing only at the $(1,1)$ element turns out to be
easily diagonalizable: 
\begin{equation*}
  \bra{n}\Hop_0\ket{m}=(1+b^2)\,n\,\delta_{nm}+
  b\,\sqrt{m(m+1)}\delta_{n,m+1}+  b\,\sqrt{n(n+1)}\delta_{m,n+1}\,.
\end{equation*}
To see this, we first change to another representation 
\begin{equation*}
  \psi_n= (-)^nb^{n}\sqrt{n}\phi_n
\end{equation*}
its action being given by 
\begin{equation*}
  \label{eq:4}
  \Hop_0 \phi_n = (1+b^2)\,n\phi_n - b^2(n+1)\phi_{n+1}-(n-1)\phi_{n-1}
\end{equation*}
\begin{lemma}
  $\Hop_0$ leaves all the subspaces $\Pop^{(k)}=\{\phi_n=\Nop n^k+
  \Oop(n^{k-1})\}$, $(k=0,1,2,...)$, invariant.
\end{lemma}
\begin{proof} By inspection, the terms $n^{k+1}$ exactly cancel and the $n^k$
term is multiplied by $(1-b^2)(k+1)$. 
\end{proof}
\begin{thm}
  $\Hop_0$ has discrete spectrum given by $E_n=|1-b^2|\,n, (n=1,2,3,...)$. 
\end{thm}
\begin{proof}
  In each finite-dimensional subspace $\Pop^{(k)}$ the matrix is
upper-triangular, hence its eigenvalues can be read off the diagonal. The
eigenvectors are given by the basis of orthogonal polynomials
w.r.t. the discrete measure
\begin{equation*}
  \vert\phi\vert^2=\sum_{n=1}^\infty n\,b^{2n}\,|\phi_n|^2
\end{equation*}
which are known as Meixner polynomials
\cite{szegoe75:_orthog_polyn}. They can be defined by
\begin{equation*}
  \phi^{(k)}_n=(1-b^2)\sqrt{\frac{k+1}{b^{2(k+1)}}}\,_2F_1(-k,n+1,2,1-b^2)
\end{equation*}
and satisfy the orthonormality property
\begin{equation*}
  \sum_{n=1}^\infty   n\,b^{2n}\phi^{(k)}_n  \phi^{(h)}_n=\delta_{k,h}
\end{equation*}
 \end{proof}

 The rank-one-perturbation property \cite{golub96} gives now the
 spectrum of $\Mop$ in the $F=0$ sector. The eigenvalue equation reads
\begin{equation*}
   1 = b^2\, \sum_{k=0}^\infty \dfrac{|\braket{v_1}{\phi^{(k)}}|^2}{(k+1)(1-b^2)-E}
\end{equation*}
where $v_1$ is the first basis vector in the representation
of~(\ref{eq:1}), and we find
\begin{equation*}
  b^2\,\sum_{k=1}^\infty \dfrac{k b^{2(k-1)}(1-b^2)^2}{(1-b^2)\,k - E} = 1
\end{equation*}
hence, by setting $E=(1-b^2)\epsilon$, one gets
\begin{equation*}
  \Delta(\epsilon)=\sum_{k=1}^\infty \dfrac{k\,b^{2k}}{\epsilon-k}
    =\frac{1}{1-b^2}
\end{equation*}
The equation can be re-expressed in terms of hypergeometric functions:
 \begin{equation*}
   \Delta(\epsilon) = 
   1+\frac{\epsilon b^2}{\epsilon-1} \, _2F_1(1,1-\epsilon,2-\epsilon,b^2)
 \end{equation*}
which is equivalent to what has been derived in
\cite{veneziano06:_planar_quant_mechan} by another method. In our
derivation the structure of $\Delta(\epsilon)$ arises naturally via
the ``rank-one-perturbation'' theorem.

\section*{Appendix 2}
We want to discuss the structure of the superleading term in
Eq.(\ref{eq:8}).  Since we work at $N=\infty$ we can study the
representation of $T_1$ in the partitions basis, which in this limit
is orthogonal. The leading behavior of the normalization factor can
be easily expressed by representing a partition in terms of
``composition'', i.e.
\begin{equation*}
  \lambda=\{1^{r_1}\,2^{r_2},\ldots k^{r_k}\}
\end{equation*}
the integers $r_j$ denoting the multiplicity of $j$ in the partition
of $n=\sum_j j\,r_j$. We find
\begin{equation*}
  \Nop_\lambda^{-2} =   \braket{\lambda}{\lambda}\approx N^{|\lambda|}\prod_jj^{r_j}\,r_j!
\end{equation*}
Now let's apply $\Tr{a^{\dagger 4}}$ to the normalized
$\ket{\lambda}$. We find
\begin{equation*}
  \Tr{a^{\dagger 4}}\ket{\lambda}=  \Tr{a^{\dagger 4}}\, \ket{r_1,r_2,\ldots,r_k}\approx \frac{\Nop_{\lambda}}{\Nop_{\lambda'}} \ket{r_1,r_2,..,r_4+1,...}=2N^2\sqrt{r_4+1}\,\ket{\lambda'}
\end{equation*}
where $\lambda'=\lambda\cup\{4\}$. The operator $\Tr{a^{\dagger 4}}$
when multiplied by $\quart b^2/N$ gives rise to one component of $T_1$
which couples states in the form
\begin{equation*}
  \lambda\to \lambda\cup\{4\} \to \lambda\cup\{4^2\}\to\lambda\cup\{4^n\}\ldots
\end{equation*}
with matrix elements  which are identical to those of a single creation
operator. The action of $\Tr{a^4}$ is simply obtained by
transposition; if a partition does not contain $2$ or $4$ it is
annihilated by $\Tr{a^4}$, which means, of course, that the resulting
matrix elements are not $O(N)$, but finite.  The analysis of the remaining
operators in $T_1$, coming from $H_2$, is much more involved, but the
structure can be identified by using computer algebra. Eventually we
get the following representation
\begin{eqnarray*}
  T_1\ket{r_1,r_2,...,r_4,...} &=& 
  \sqrt2\,b^2\,N\,(\sqrt{r_2}\;\ket{r_1,r_2-1,...,r_4,...} + 
  \sqrt{(r_2+1)}\;\ket{r_1,r_2+1,...,r_4,...})\\
  &+&    \half b^2\,N\,(\sqrt{r_4}\;\ket{r_1,r_2,...,r_4-1,...} + 
  \sqrt{r_4+1}\;\ket{r_1,r_2,...,r_4+1,...})
\end{eqnarray*}
Notice that in this way we can identify an infinite sequence of
subspaces invariant under the action of $T_1$: each subspace is
labeled by a partition characterized by $r_2=r_4=0$. This is a
``vacuum'' state for $T_1$. The action of $T_1$ builds a subspace
isomorphic to the Hilbert space of a two-dimensional harmonic
oscillator and the action of $T_1$ in all subspaces is always the
same. The harmonic oscillator frequencies can be read off the diagonal
part $T_0$, namely $(\omega_1,\omega_2)=(2,4)(1+b^2)$. Hence we find
that indeed the only effect of $T_1$ at this level consists in a shift
of the whole spectrum proportional to $b^2\,N$ exactly as it happens
to a simple harmonic oscillator under perturbation by a term $\propto
(a+a^\dagger)$.

\bibliographystyle{amsalpha}

\providecommand{\bysame}{\leavevmode\hbox to3em{\hrulefill}\thinspace}
\providecommand{\MR}{\relax\ifhmode\unskip\space\fi MR }
\providecommand{\MRhref}[2]{%
  \href{http://www.ams.org/mathscinet-getitem?mr=#1}{#2}
}
\providecommand{\href}[2]{#2}

\end{document}